\documentclass[12pt,preprint]{aastex}
\usepackage{natbib}
\bibpunct{(}{)}{;}{a}{}{,} 

\begin{document}

\title{Which Radial Velocity Exoplanets Have Undetected Outer Companions?}
\author{Timothy J. Rodigas and Philip M. Hinz}
\affil{Steward Observatory, University of Arizona, 933 N Cherry Ave. Tucson AZ 85721-0065; email: rodigas@as.arizona.edu}

\begin{abstract}
The observed radial velocity (RV) eccentricity distribution for extrasolar planets in single-planet systems shows that a significant fraction of planets are eccentric ($e > 0.1$). However an RV planet's eccentricity, which comes from the Keplerian fitting, can be biased by low signal-to-noise and poor sampling. Here we investigate the effects on eccentricity produced by undetected outer companions. We have carried out Monte Carlo simulations of mock RV data to understand this effect and predict its impact on the observed distribution. We first quantify the statistical bias of known RV planets' eccentricities produced by undetected zero-eccentricity wide-separation companions and show that this effect alone cannot explain the observed distribution. We then modify the simulations to consist of two populations, one of zero-eccentricity planets in double-planet systems and the other of single planets drawn from an eccentric distribution. Our simulations show that a good fit to the observed distribution is obtained with 45$\%$ zero-eccentricity double-planets and 55$\%$ single eccentric planets. Assuming our two simulated populations of planets are a good approximation for the true RV population, matching the observed distribution allows us to determine the probability that a known RV planet's orbital eccentricity has been biased by an undetected wide-separation companion. Averaged over eccentricity we calculate this probability to be $\sim 4\%$, suggesting that a small fraction of systems may have a yet to be discovered outer companion. Our simulations show that moderately-eccentric planets, with $0.1 < e < 0.3$ and $0.1 < e < 0.2$, have a $\sim 13\%$ and $\sim 19\%$ probability, respectively, of having an undetected outer companion. We encourage both high-contrast direct imaging and RV follow-up surveys of known RV planets with moderate eccentricities to test our predictions and look for previously undetected outer companions.
\end{abstract}


\section{Introduction}
Since the 1990s, over 300 extrasolar planets have been discovered (http://exoplanet.eu). Exoplanets have been discovered by several different techniques, namely radial velocity (RV), transits, astrometry, microlensing, and most recently direct imaging (e.g., \cite{kalas,marois}). The RV method, which has discovered the majority, monitors the periodic velocity shifts (in the radial direction) of a given star. These velocity variations are caused by the star orbiting the barycenter of the combined system, which can contain additional companions (typically stars, brown dwarfs, and/or planets). The RV method allows one to extract the orbital period of the companion ($P$), its semi-major axis $a$, the lower limit to its mass ($m \sin i$), and its eccentricity ($e$). Since there are $\sim$ 300 RV-discovered exoplanets, we have a large enough sample to construct distribution functions for each of these quantities.

Of these distributions, the eccentricity distribution is the most puzzling. In our own solar system, planets have very low eccentricity values (nearly circular orbits). Jupiter, the gas giant to which we compare discovered extrasolar planets, orbits with $e = 0.05$. While the RV extrasolar planet eccentricity distribution has a strong peak near $e = 0$, it also has a significant tail extending all the way out to $e = 0.93$ (see Fig. \ref{eccdist}). If our solar system and its planets are common and ordinary, then we would expect most exoplanets to have low eccentricities. Clearly this is not the case.    

There have been many attempts to explain the observed eccentricity distribution. Most of these focus on planet-planet interactions and scattering (e.g. \cite{rasio,levison,adams,juric}). These interactions can build up the eccentricities of two or more planets, resulting in scattering or ejection. In the process, the remaining planets settle down into stable, high-eccentricity orbits. While scattering is certainly plausible and can match the $e \gtrsim 0.2$ observed eccentricity distribution, it is unclear whether the majority of planetary systems go through such violent planet-planet interactions.

Other factors may be at play. \cite{shen} explored the effects of RV systematics on eccentricity. They showed that planets' eccentricities can increase with lower signal-to-noise ratios (SNR), given by $K/\sigma$, where $K$ is the velocity semi-amplitude and $\sigma$ is the experimental error (typically depending on the telescope and instrument error as well as stellar jitter). Specifically $e$ increases for $K/\sigma \lesssim 3$. They also show that eccentricity increases with fewer observations in a given RV data set ($N_{obs} \lesssim 60$). Here, the number of observations is the number of data points in a star's radial velocity curve. Since the SNR and $N_{obs}$ conditions are not always met for RV data, especially for the earliest discoveries, we can expect that this bias plays a role in the observed eccentricity distribution. \cite{shen} estimated that about 10$\%$ of RV-detected exoplanets are affected by this bias.

Undetected additional planets in planetary systems can also affect RV eccentricities. When two planets orbit a star, both contribute to the star's radial velocity. Assuming the planets are not gravitationally interacting with each other, one produces the total radial velocity curve by adding the radial velocity data from each planet. One can detect the second planet by monitoring the star long enough to observe its full period. Since this time frame is usually $\gtrsim 10$ years, only $\sim 20$ multiple-planet systems have been discovered. If one observes a double-planet system for a duration shorter than the period of the second planet, a long-term trend can appear in the data. \cite{fischer} explored the possible bias introduced by \textit{wide-separation planets} by injecting real long-term trend data into some of their existing one-planet RV data. After fitting the new data, they found that the mass and semi-major axis of the original planet were mostly unchanged, but that its eccentricity value  increased, sometimes by as much as $\Delta e = 0.25$. They were able to detect the long-term trends only in cases where there was dense phase coverage (large $N_{obs}$) and high SNR. These results are very important for several reasons: they confirm the effects of the systematic biases discussed by \cite{shen}, and they show that \textit{undetected} additional companions can drive up a known planet's eccentricity value. Nonetheless, we cannot draw large-scale conclusions since the \cite{fischer} results were based on a small sample of tests, and the test parameters were not significantly (randomly) varied. 

To determine the full effect of undetected wide-separation planets on eccentricity, we have carried out Monte Carlo simulations of mock RV data. We first determine the statistical effect of undetected wide-separation companions on eccentricity. We then try to match the observed eccentricity distribution with a population of zero-eccentricity planets in double-planet systems. We show that this population alone cannot describe the observed eccentricity distribution, and we therefore need a second input population of high-eccentricity planets. After matching the observed distribution, we derive the probability that a known planet has a wide-separation companion, information potentially important for both RV follow-up surveys and direct imaging. In $\S$2, we describe the simulations, test parameters, and methodology. In $\S$3, we present the results and discuss the implications. In $\S$4, we summarize and conclude. 

\section{Simulation Parameters and Methodology}
An RV-detected planet's orbital parameters ($m \sin i$, $a$, and $e$) are determined from the Keplerian fit to the star's radial velocity data. The equations needed to calculate the star's radial velocity are: 
\begin{equation}
V_{rad} = V_{0} + K~[\cos(\omega + T) + \cos(\omega)]
\end{equation} 
\begin{equation}
K = \sqrt{G/a/(M_{*} + m)}~m \sin i~/~\sqrt{1-e^{2}}
\end{equation}
\begin{equation}
T = 2~\tan^{-1}[\tan(E / 2) \sqrt{(1 + e) / (1 - e)}]
\end{equation}
\begin{equation}
E = M + e \sin E
\end{equation}
\begin{equation}
M = (t - t_{0}) \frac{2 \pi}{P}.
\end{equation}
$P$ is the orbital period of the planet derived from Kepler's laws, $t$ is the time of the observation, $t_{0}$ is the time of perihelion, and $M$ is the mean anomaly. $E$ is the eccentric anomaly and since it is given in a transcendental equation it cannot be solved analytically. For our simulations, we solve for $E$ using iterative loops. $T$ is the true anomaly, $K$ is the velocity semi-amplitude, and $\omega$ is the argument of the perihelion. We have set $\sin i = 1$, assuming edge-on systems in all cases. $V_{0}$ is the velocity offset which for simplicity we have set to zero. $M_{*}$ and $m$ are the mass of the star and planet, respectively. Since most RV stars are solar-type stars, we set $M_{*} = M_{\odot}$. 

Our simulations, aside from control simulations, consist of 100 sets of $N$ RV curves. $N = 227$ since, as of January 2009, there are 227 single-planet RV systems (http://exoplanet.eu). Thus we simulate $\sim 10^{4}$ total planets. We do not simulate the total number of observed RV systems because some of these ($\sim 30$ systems) have multiple planets. Our hypothesis applies only to single-planet RV detections. 

For each simulated planet, we draw $\omega$ randomly from a uniform distribution in the range [0,2$\pi$] and $t_{0}$ in the range [0,$P$], where $P$ is the period of that particular planet. Each planet's semi-major axis $a$ and mass $m$ are drawn randomly from the most current RV distribution functions, irregardless of being in a single or double-planet system. From the \cite{butler} catalog of exoplanets, these are $dN/dM \propto M^{-1.1}$ and $dN/d\log a \propto a^{0.4}$. In simulated systems, the first planet has $a_{inner} \in [0.05,3]$ AU and the second has $a_{outer} \in [7,15]$ AU. The first planet's orbital range was chosen because most RV-detected planets are found within 3 AU, and the closest planets have periods on the order of a few days (= 0.05 AU). We acknowledge that the \cite{butler} minimum semi-major axis used to derive the distribution function was $a \sim 0.1$, but this difference is insignificant. It is also true that RV has found some planets beyond 3 AU, but the statistics are poor for this small sample. The second planet's orbital range was chosen for several reasons: gas giants in our solar system (Jupiter, Saturn) orbit in this range, and we wanted to avoid any resonances $<$ 2:1 that might, via secular interactions, lead to higher eccentricity values. In fact, the smallest resonance we allow is $\sim 3.5:1$ (outer planet at 7 AU, inner planet at 3 AU). Our simulations are testing stable relaxed planetary systems, and the eccentricity bias we are investigating is not from planet-planet interactions but rather from the Keplerian fitting. 

The mass range for the first planet was chosen to be $m_{inner} \in [0.01,25]~M_{J}$ ($M_{J}$ is the mass of Jupiter). These limits reflect the $\sim$ minimum and maximum mass planets discovered by RV so far (http://exoplanet.eu). The mass range for the second planet was chosen to be $m_{outer} \in [m_{inner}/M_{J},25]~M_{J}$. If a planetary system has two planets, then the outer planet is probably at least as massive as the inner. Though speculative, there is some evidence for this in \cite{wright}, who show that there is on average a 1-1 correspondence in mass for the RV-discovered two-planet systems. It is possible that this result is biased by the fact that more distant planets have to be more massive to be detected by the RV method. Nonetheless most multiple-planet systems have been discovered by RV, so the sample in \cite{wright} is the largest and most statistically-significant sample from which to draw conclusions. 

\begin{figure}[t]
\centering
\includegraphics{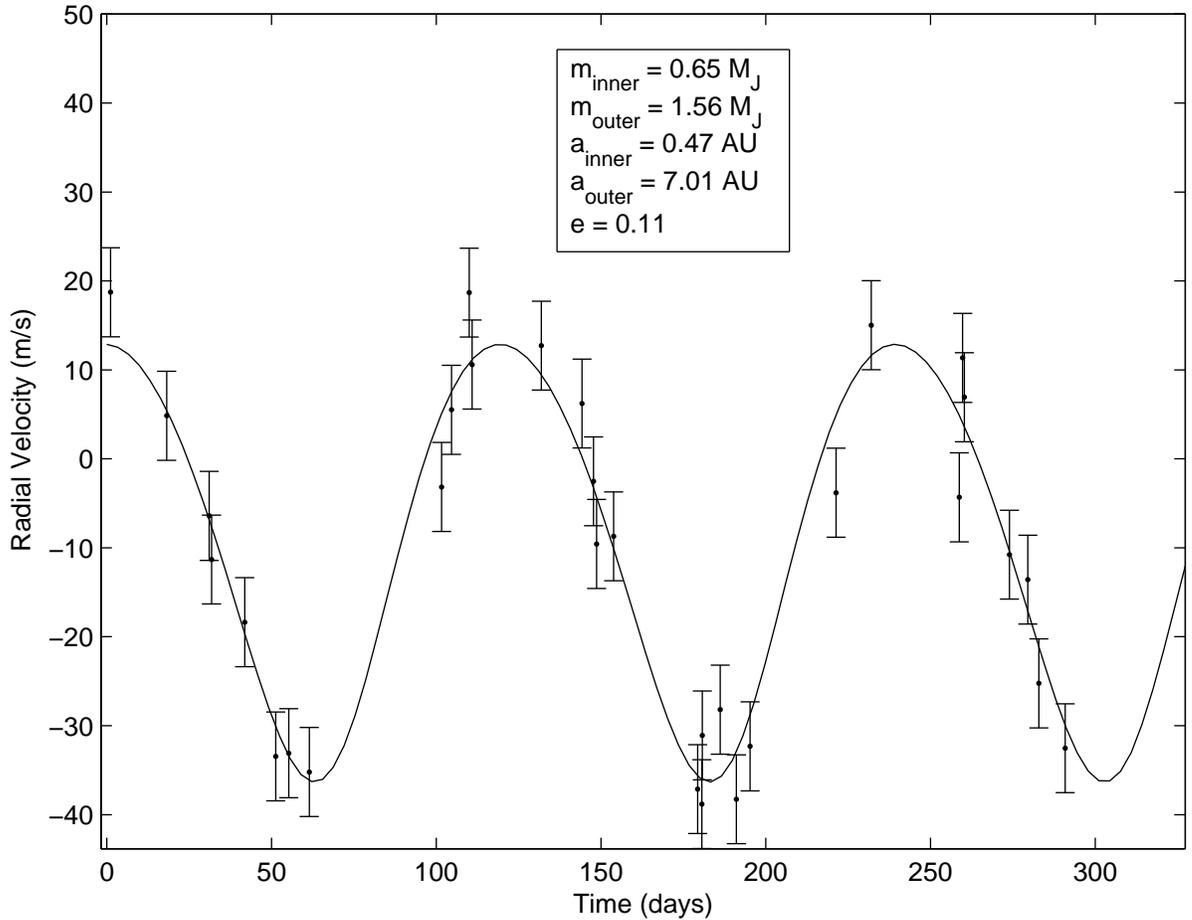}
\caption{Typical radial velocity plot from the simulations. In this case there are two planets in the system, both contributing to the star's radial velocity. There is no apparent long-term trend from the second planet evident in the data, but the fitted eccentricity has increased from $e = 0$ to $e = 0.11$.}
\label{radvel}
\end{figure}

For each simulation, we compute the star's radial velocity $V_{rad}$ at 32 points in time over 2$P$. In all cases (one-planet system or two-planet system), $P$ is the period of the inner planet. We have chosen 32 data points because this is a moderate amount of observations. We acknowledge that since $N_{obs} = 32 < 60$, eccentricities will be slightly biased (as described by \cite{shen}). Nonetheless, $N_{obs} = 32$ is a reasonable assumption given the typical $N_{obs}$ for an actual RV data set. We have chosen to simulate each system's data for two periods because most RV data is published with at least two periods of coverage. We emphasize that the 32 data points along each RV curve are not equally spaced. This would not be realistic. Instead we have tried to mimic the ``clustering" of data points seen in typical RV data (see Fig. \ref{radvel}). To accomplish this, we divide a given simulated data set into 8 zones, each containing 4 points (observations). In each zone, the first point's location is determined randomly. The subsequent points' locations are also determined randomly, but the inner boundary for each is the location of the previous point. This causes an apparent grouping, or clustering, of observations in each zone.

If there are two planets in a simulated system, we compute the radial velocity independently for each planet and then add them together to get the total $V_{rad}$. In all cases, we add Gaussian noise with $\sigma = 5$ m/s to $V_{rad}$. This $\sigma$ was chosen as a good estimate of typical RV precision. Because we enforce $K > 3 \sigma$, changing $\sigma$ to a higher (potentially more RV-representative) value (say, 10 m/s) would have little to no effect on our results. All simulations were carried out in Matlab. To determine the orbital parameters, we cycle through $e$ from 0 to 0.99, hold $e$ constant at each value, and fit for the other orbital parameters. We then take the lowest $\chi^{2}$ solution, along with its corresponding $e$, as the final set of orbital parameters. Because we know the input parameters for each simulated system beforehand, we do not use periodograms and false-alarm probability tests in our fitting procedure. When fitting a data set, we only fit for \textit{one} planet because we are testing the effects of \textit{undetected} wide-separation planets on eccentricities. To ensure that the outer planets do not create long-term trends in the data, we enforce $\chi_{reduced}^{2} < 3$. This limiting value was chosen because RV data with poorly-constrained fits are rarely published; if they are, the authors typically suggest that a long-term trend is present and attempt a two-planet Keplerian fit.   

\section{Results and Discussion}
\subsection{Control Simulations}
Before running our full suite of simulations, we checked that our Keplerian fitting was working correctly. To test this, we simulated 50 sets of $N = 227$ single-planet systems ($\sim 10,000$ planets). Each planet's semi-major axis and mass were drawn from the appropriate distributions. We set $e$ to zero, which allows us to directly measure any unforeseen effects on eccentricity. If the fit is working correctly, then the output parameters for each planet should match the input parameters within experimental error.  

\begin{figure}[t]
\centering
\includegraphics{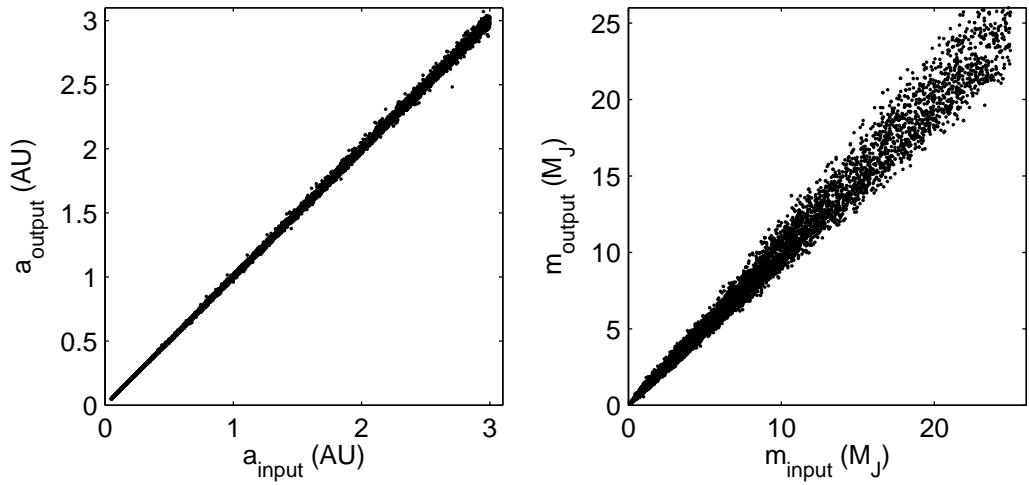}
\caption{\textit{Left}: output semi-major axis as a function of the input semi-major axis for the control simulation (one-planet systems). \textit{Right}: the same, except showing the relationship between output and input mass. In each case, the tight linear relationships and slopes of unity verify that our simulations have no unforeseen effects on a given planet's semi-major axis or mass.}
\label{control1p}
\end{figure}

\begin{figure}[t]
\centering
\includegraphics{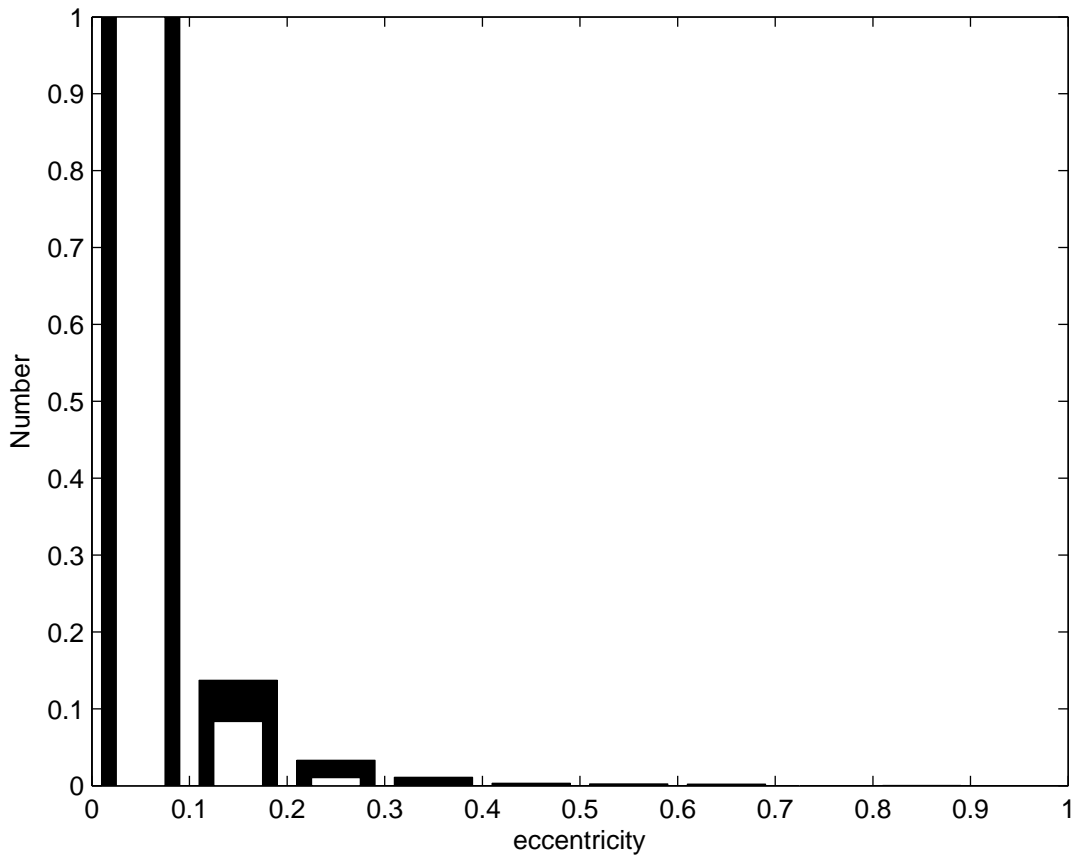}
\caption{Eccentricity distribution for the control simulation (one-planet systems) (white bars), and the eccentricity distribution for simulated two-planet systems (black bars), both normalized to the respective first bin. All planets had their input eccentricities set to zero. For one-planet systems, $8\%$ of the simulated planets' fitted eccentricities have increased, most likely due to biases coming from low SNR and poor sampling (discussed in \cite{shen}). When the wide-separation companions are introduced, the fraction of eccentric ($e > 0.1$) planets increases by $\sim 10\%$ to $\sim 18\%$.}
\label{esimdist}
\end{figure}

Fig. \ref{control1p} shows the relationships between $a_{output}$ and $a_{input}$, and $m_{output}$ and $m_{input}$. The tight linear relationships and slopes of unity show that our Keplerian fitting has no effect or bias on a planet's semi-major axis or mass. Fig. \ref{esimdist} shows the average distribution of fitted eccentricity values, normalized to the first bin. Fig. \ref{esimdist} reveals that $\sim 8\%$ of planets are eccentric ($e > 0.1$). This small bias is due to the effects discussed by \cite{shen}. Namely, our $N_{obs} = 32 < 60$, which is the suggested number of observations. Furthermore, although \cite{shen} did not explore this, we believe that the ``clustering" of data points also affects a planet's eccentricity. This is a typical feature of published RV data, so we have chosen not to ignore it.   

\begin{figure}[t]
\centering
\includegraphics{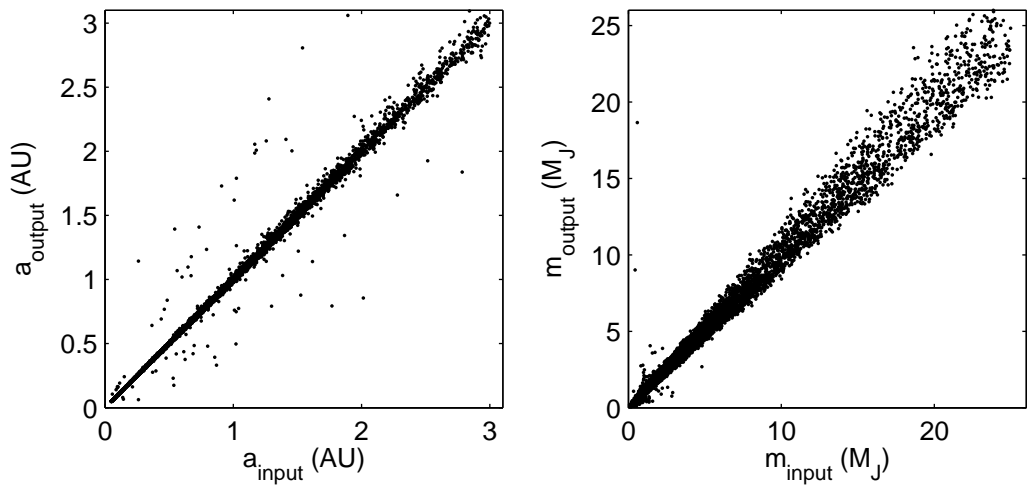}
\caption{\textit{Left}: output semi-major axis as a function of the input semi-major axis in simulated systems with two planets, both with zero eccentricity. \textit{Right}: the same, except showing the relationship between output and input mass. In both cases, some outliers have been clipped. The tight linear relationships and slopes of unity show that wide-separation companions have a negligible effect on the fitted semi-major axis and mass, which agrees with expectations and the previous work by \cite{fischer}. It is worth noting, however, the increased scatter in output semi-major axis as compared to output mass. This suggests that semi-major axis is the more easily affected quantity.}
\label{control2p}
\end{figure}

Having confirmed the validity of our Keplerian fitting routine, the next step was to quantify the statistical effects of \textit{undetected} wide-separation planets on one-planet fits. To this end, we simulated 100 sets of $N \simeq 100$ two-planet systems ($\sim 10,000$ planetary systems). Each planet's semi-major axis and mass were drawn from the appropriate distributions, and we again fixed $e$ at zero. Comparing the results of these double-planet simulations to those of the single-planet simulations yields the second planet's effects on the first planet's parameters. Fig. \ref{control2p} shows the relationships between $a_{output}$ and $a_{input}$, and $m_{output}$ and $m_{input}$. In all cases, $a$ and $m$ are the semi-major axis and mass of the \textit{inner} planet. Fig. \ref{control2p} shows tight linear relationships with slopes of unity. Evidently wide-separation planets have no effect on the inner planet's mass and semi-major axis, although there is certainly more scatter. These results agree with the previous work by \cite{fischer}.

Fig. \ref{esimdist} also shows the average distribution of fitted eccentricity values for two-planet systems. The percentage of planets with $0.1 < e < 0.2$ is $\sim 13\%$, and the total percentage of planets with $e > 0.1$ is $\sim 18\%$. Thus the addition of the second planet can increase the eccentricity of the first planet. Knowing that the $N_{obs} < 60$ effect accounts for $\sim 8\%$ of the biased eccentricities, the sole effect from wide-separation planets is $\sim 10\%$. We note that ``wide-separation" means $7<a_{outer}/$AU~$<$~15, the final range we settled on. We have also tested other semi-major axis ranges of different extents, such as [10,30] AU. In general, we saw that the larger the semi-major axis range, the lower the eccentricity bias. This makes sense given the RV equations (Eqs. 1-5); closer planets increase the velocity semi-amplitude $K$ which in turn affects $V_{rad}$ and the parameters of the fitted planet. In our simulations, however, bringing the second planet closer in does not necessarily induce more eccentricity bias. This is because the Keplerian fit becomes poor and the system is rejected. Therefore there is a ``sweet spot" ($\sim 7$ AU) where eccentricity is slightly biased, no long-term trend appears in the data, and the Keplerian fit converges.

\begin{figure}[t]
\centering
\includegraphics{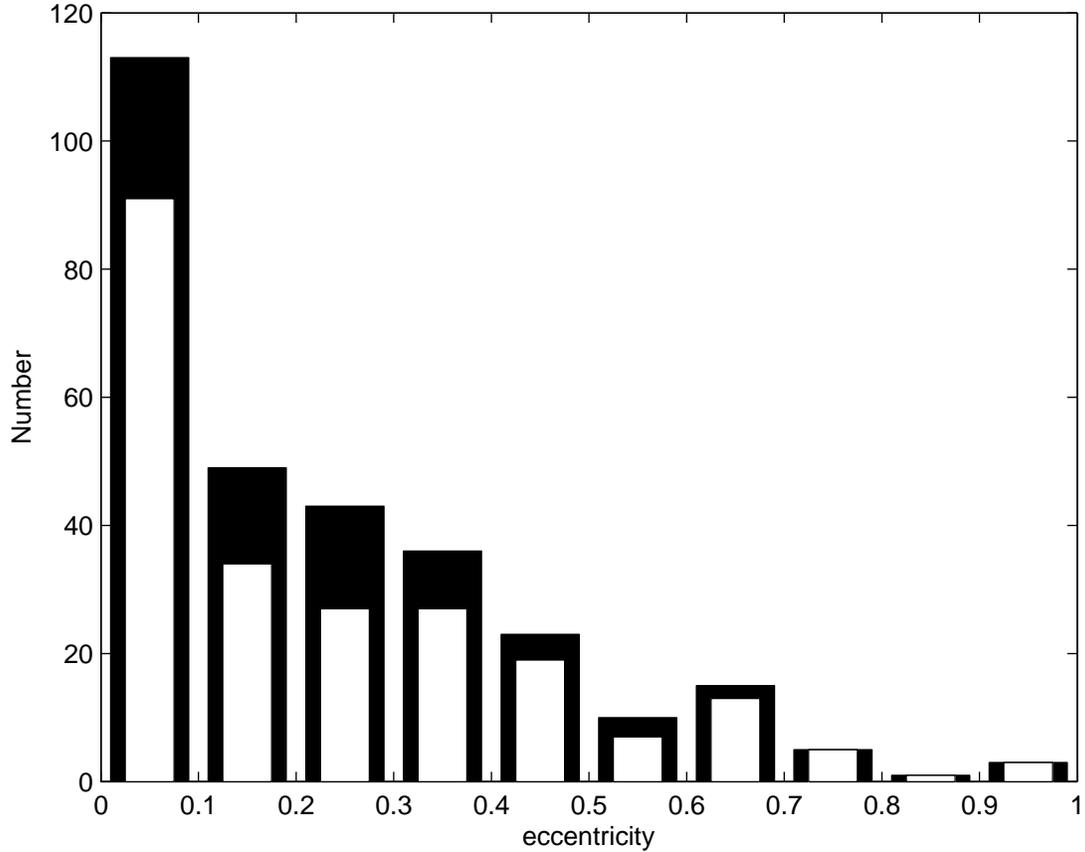}
\caption{The observed RV eccentricity distribution (black bars) and the observed \textit{single-planet} RV eccentricity distribution (white bars), both from http://exoplanet.eu as of January 2009. We compare our simulations only to the single-planet distribution because we hypothesize that undetected outer companions bias \textit{one-planet} Keplerian fits. Note the large number of high eccentricity ($e > 0.1$) planets.}
\label{eccdist}
\end{figure}

The double-planet eccentricity distribution in Fig. \ref{esimdist} led us to draw two additional conclusions. First, our results agree with those of \cite{fischer}. The particular scatter in eccentricity values ($e$ as high as $\sim 0.3$) matches the maximum $\Delta e$ they reported. If we relaxed our fitting constraints, then we would surely see higher eccentricity values. Therefore our agreement with the \cite{fischer} result suggests that our fitting constraints (in particular $\chi_{reduced}^{2} < 3$) are acceptable. Second, we cannot match the \textit{observed} single-planet RV eccentricity distribution (Fig. \ref{eccdist}) with \textit{only} zero-eccentricity double-planet systems. Fig. \ref{esimdist} at best only matches the $e < 0.2$ observed distribution shown in Fig. \ref{eccdist}. Consequently we can reject the notion that the observed distribution can be explained by 100$\%$ zero-eccentricity double-planet systems. This result has two interesting implications. It lends more support to the growing evidence that the true RV population consists of a significant fraction of high-eccentricity planets in single-planet systems; and, following from this conclusion, our solar system, as current evidence continues to show, is probably not ordinary. This is to say that the number of planets in our solar system and the low eccentricity of each is uncommon. If we had been able to match the observed eccentricity distribution with 227 zero-eccentricity double-planet systems, we could suggest that multiple-planet systems with low-eccentricity planets (like our own) may be more common than previously believed. However our simulations show that multiple-planet systems cannot be the entire story.

\subsection{Matching the Observed Eccentricity Distribution}
Our preliminary results show that we cannot match the observed single-planet eccentricity distribution with only zero-eccentricity double-planet systems. High-eccentricity planets in single-planet systems must constitute a significant fraction of the true RV population. Therefore, in addition to a simulated population of zero-eccentricity double-planets, we introduce a second population of single truly-eccentric planets. When simulating a truly-eccentric single planet, we draw its eccentricity value randomly from the Schwarzschild distribution for eccentricity $e$, given by
\begin{equation}
dN/de = \frac{e}{\sigma_{e}^2} \exp(\frac{-e^2}{2 \sigma_{e}^2}),
\end{equation}
with $\sigma_{e} = 0.3$. This is the distribution in eccentricity one would expect to see from gravitational scattering of astronomical bodies, whether they are stars or planets. Furthermore, \cite{juric} used it to generate eccentricities in their simulations, and it matches the $e \gtrsim 0.2$ observed eccentricity distribution very well. 

\begin{figure}[t]
\centering
\includegraphics{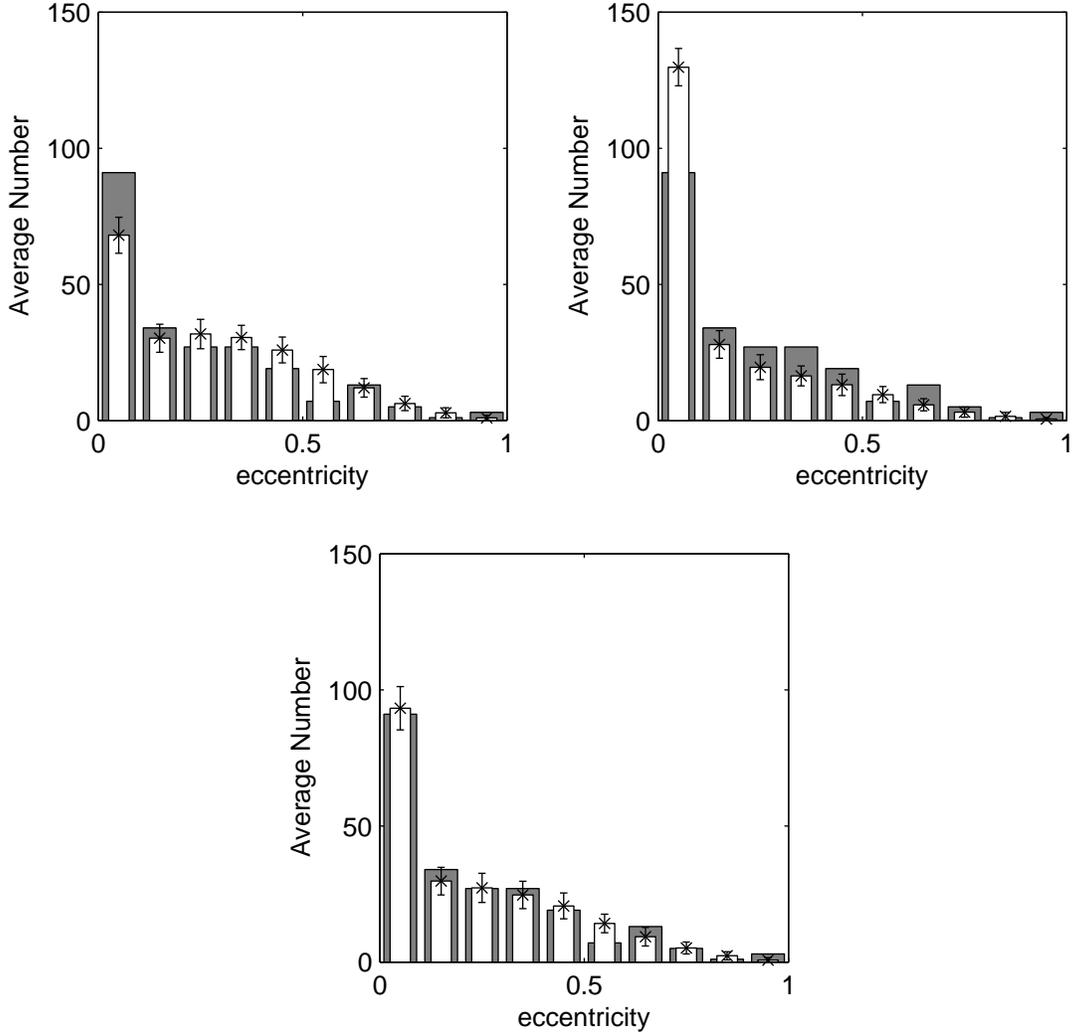}
\caption{Simulated eccentricity distributions (skinny white bars with markers and error bars in all three plots) and the observed single-planet eccentricity distribution (wide grey bars in all three plots) from http://exoplanet.eu. \textit{Top-left}: simulated distribution for R = 30$\%$. The number of low-eccentricity planets is underestimated and the number of high-eccentricity planets is overestimated. \textit{Top-right}: simulated distribution for R = 65$\%$. This time the number of low-eccentricity planets is overestimated and the number of high-eccentricity planets is underestimated. \textit{Bottom}: simulated distribution for R = 45$\%$. This distribution matches the observed single-planet eccentricity distribution very well ($\chi_{reduced}^{2} = 1.74$, disagreeing only around $e \sim 0.55$ where the observed distribution dips down, most-likely due to small-number statistics).}
\label{matchdist}
\end{figure}

\begin{figure}[t]
\centering
\includegraphics{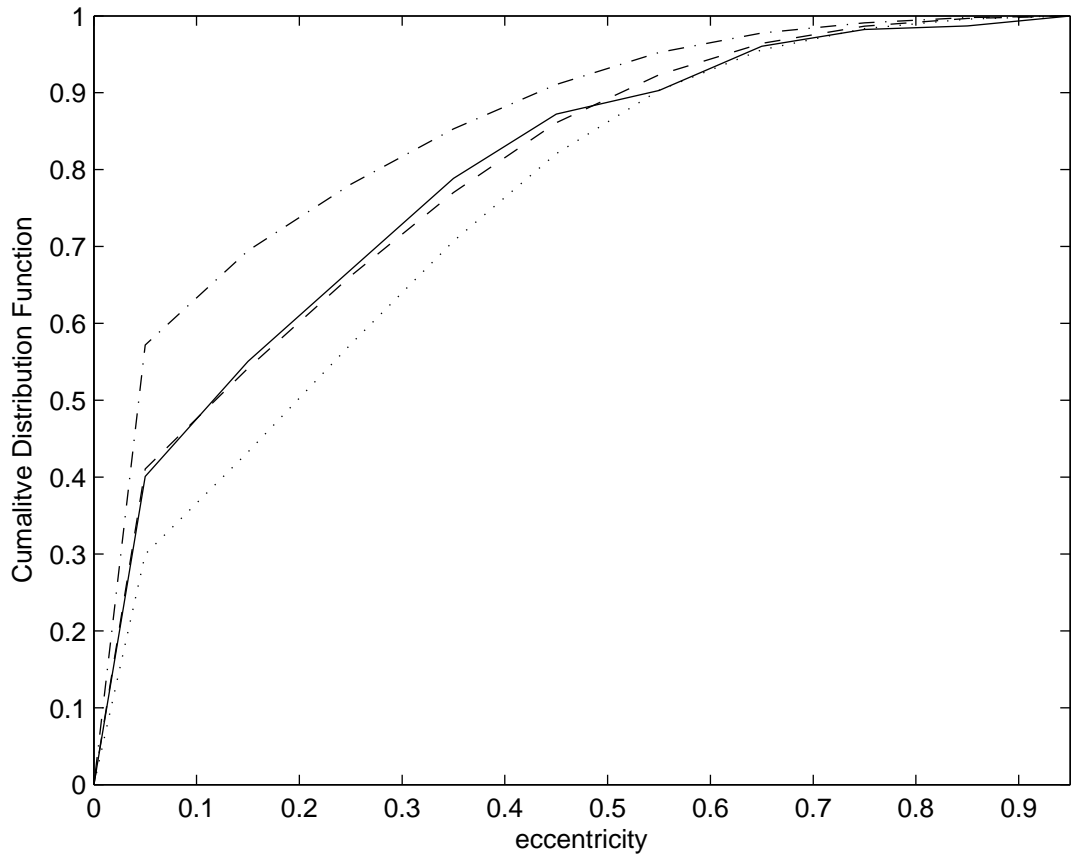}
\caption{Cumulative distribution functions of the simulated eccentricity distributions for R = 30$\%$ (dotted line), 45$\%$ (dashed line), 65$\%$ (dot-dashed line), and the observed single-planet eccentricity distribution (solid line). The R = 45$\%$ curve matches the observed, but the R = 30$\%$ and R = 65$\%$ do not agree.}
\label{cdf}
\end{figure}

Fig. \ref{matchdist} shows the results of our final simulations, comparing three simulated distributions to the observed single-planet eccentricity distribution from http://exoplanet.eu. The ratio of double-planet systems to total systems ($\equiv R$) varies for each of the different simulations between $30\%, 45\%,$ and $65\%$. Fig. \ref{matchdist} shows that the $R = 30\%$ case underestimates the number of low-eccentricity planets and overestimates the number of high-eccentricity planets. The $R = 65\%$ case shows the opposite. The $R = 45\%$ distribution is visibly the best match, with each point at or within 1 sigma of the observed distribution. This case also had the lowest $\chi_{reduced}^{2}$ value (= 1.74) compared to the other two distributions, which had $\chi_{reduced}^{2} > 2$. Fig. \ref{cdf} shows the cumulative distribution functions of the simulated and observed distributions. Again the overall agreement is evident for the $R = 45\%$ distribution. We also performed a two-sample K-S test for the $R = 45\%$ case. The null hypothesis was that this simulated distribution and the observed distribution from http://exoplanet.eu were the same. The two-sample K-S test showed that the null hypothesis could not be rejected at the $\alpha = 0.05$ significance level (95$\%$ confidence), supporting the notion that the $R = 45\%$ distribution matches the observed distribution very well. These statistical results, as well as the final simulation parameters, are detailed in Table 1. 

To zero in on the optimal ratio $R$, we carried out several simulations that varied $R$ around 45$\%$. Together with the $R = 30\%$ and 65$\%$ simulated distributions, we were able to quadratically fit $\chi_{reduced}^{2}$ as a function of $R$. This yielded a minimum $\chi_{reduced}^{2}$ of 1.6, corresponding to $R_{min} = 42_{-8.7}^{+8.2}\%$ (1-sigma limits). 

\begin{figure}[t]
\centering
\includegraphics{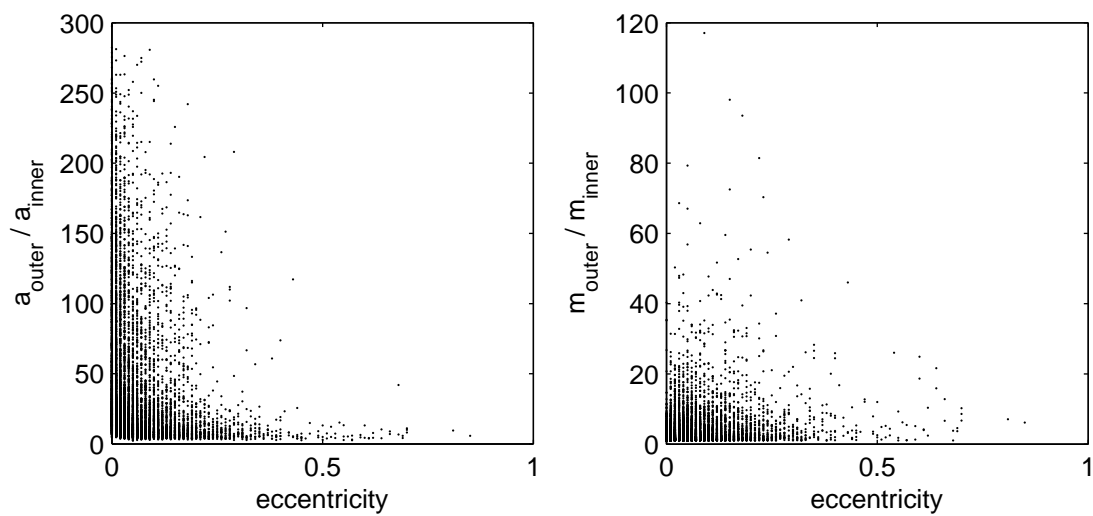}
\caption{\textit{Left}: ratio of the semi-major axes of the two planets in each simulated two-planet system as a function of fitted eccentricity. \textit{Right}: the same, except the ratio of the masses. The semi-major axis ratio increases for low eccentricity and decreases for high eccentricity, but there is no apparent correlation between mass ratio and eccentricity. This suggests that fitted eccentricity is much more dependent on the separation between two planets than on the ratio of their masses.}
\label{ratios}
\end{figure}

In a two-planet system, what determines if the fitted planet's eccentricity is biased? We already know the effect depends on the semi-major axis and mass of the second planet from Eqs. 1-5, but which of these is the dominating factor? Fig. \ref{ratios} shows the ratio of the semi-major axes and the ratio of the masses of the two planets in each double system as a function of fitted eccentricity. From Fig. \ref{ratios} it is evident that a lower $e$ means a larger separation between the two planets, and a higher $e$ corresponds to a smaller separation. There is no apparent correlation in mass with eccentricity. This tells us that in a one-planet Keplerian fit, the fitted eccentricity value depends much more on how far away the second planet is than on how massive it is.  

\subsection{Eccentricity as an Indicator of Multiplicity}
\begin{figure}[t]
\centering
\includegraphics{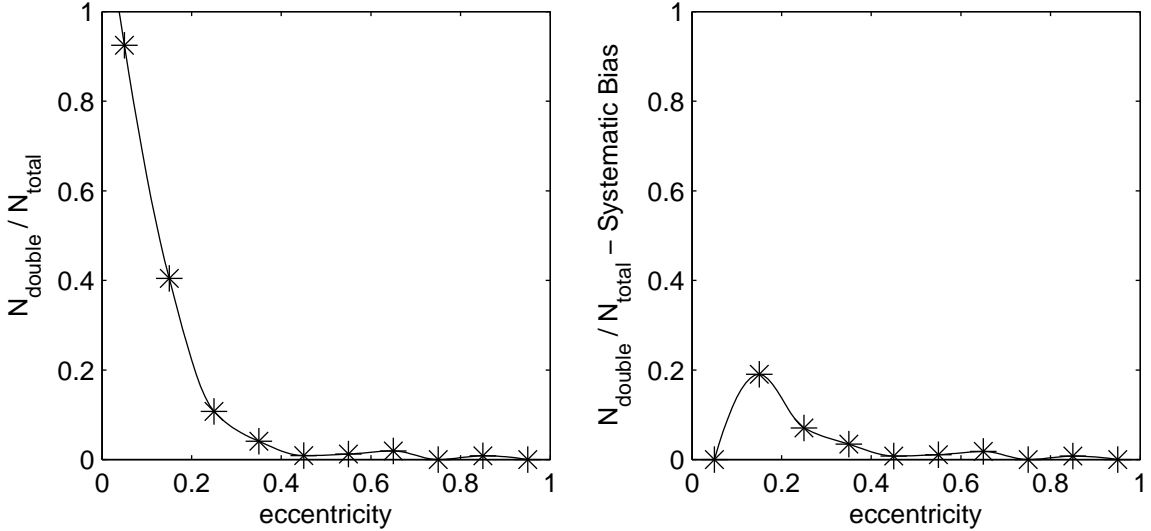}
\caption{\textit{Left}: ratio of double-planet systems to total systems as a function of fitted eccentricity. We cannot take these results at face value because the RV systematic biases (low SNR, poor sampling) have not been accounted for and subtracted off. \textit{Right}: probability that a known RV planet's eccentricity has been biased by an undetected wide-separation companion. To produce this plot, the control simulation bias (see Fig. \ref{esimdist}) has been normalized to the first bin and subtracted off, explaining the first bin's zero probability. In actuality planets with $0 < e < 0.1$ have some small non-zero probability, but the degeneracy in eccentricity (single-planet system with $e \sim 0$ or double-planet system with $e \sim 0$ and an extremely distant companion) prevents us from drawing any conclusions about double-planet probability. For $e > 0.1$, there is a non-zero probability that a given RV planet has a wide-separation companion. Specifically, for $0.1 < e < 0.4$ the average probability is $\sim 10\%$, for $0.1 < e < 0.3$ the average probability is $\sim 13\%$, and for $0.1 < e < 0.2$ the probability is $\sim 19\%$. Averaged over all eccentricities, the probability is $\sim 4\%$.}
\label{probability}
\end{figure}

The ratio $R$ tells us the fraction of simulated systems that have wide-separation companions, but we do not know which known RV systems might be affected by outer-planet bias. Because we have succeeded in matching the observed eccentricity distribution, we can use eccentricity as an indicator of exoplanet multiplicity. We stress that the discussion hereafter assumes that our two simulated input distributions of planets are a good approximation for the true RV population. To determine what information about multiplicity a planet's eccentricity yields, we calculate the probability that an RV-detected planet's eccentricity has been biased by a wide-separation companion. Fig. \ref{probability} plots the ratio of the average number of double-planet systems to total systems as a function of fitted eccentricity. The first bin ($0 < e < 0.1$) has a possible degeneracy in eccentricity: a planet with an eccentricity in this range either lives alone or has an extremely distant companion. In both cases, the eccentricity value is low and unbiased. We have therefore set the first bin to zero by subtracting off the control simulation bias (see Fig. \ref{esimdist}). In this way any unforeseen bias introduced in our simulations would not corrupt our probabilities. 

From Fig. \ref{probability}, the probability that a given planet with any eccentricity ($0 < e < 1$) has an undetected outer companion is $\sim 4\%$. For $e \geq 0.1$, any wide-separation companions have small enough orbital separations to induce effects on fitted eccentricity. For $0.1 < e < 0.4$, the average probability is $\sim 10\%$. This value increases to 13$\%$ and 19$\%$ if we consider planets with $0.1 < e < 0.3$ and $0.1 < e < 0.2$, respectively. 

Thus we now have a specific range in eccentricity values that can be used for observation target selection. From http://exoplanet.eu, there are 84 planets (in single-planet systems) with $0.1 < e < 0.4$, 57 planets with $0.1 < e < 0.3$, and 30 planets with $0.1 < e < 0.2$. Therefore, with the corollary stated above, we estimate that about 16 known RV planets should have wide-separation companions. We encourage high-contrast direct imaging observations of these targets to look for potential wide-separation companions. For a small target sample size, we suggest that direct imaging surveys focus on RV planets with $0.1 < e < 0.2$ or $0.1 < e < 0.3$ where the probabilities are highest. In this case, our simulations predict that between 6-10 RV planets should have wide-separation companions, depending on which eccentricity range is used.  

Our results are important for RV follow-up surveys as well. We again suggest looking at planets with $0.1 < e < 0.3$. For the zero-eccentricity planets, there is certainly the possibility that they are not part of multiple-planet systems, in which case RV follow-up would yield null-results. If these planets do have wide-separation companions, they are probably so far away that RV follow-up would require extremely long observation time frames to detect anything. However for $e > 0.1$, RV follow-up could reveal long-term trends and potentially extract the periods of wide-separation companions, since our simulations predict that such planets would not be too distant relative to the inner planets (see Fig. \ref{ratios}). We summarize the suggested observational strategy for RV follow-up and direct imaging in Table 2. 

\subsection{Why Not Observe Zero-Eccentricity RV Planets?}
One might wonder why we do not advocate direct imaging of known zero-eccentricity RV planets. After all, Fig. \ref{ratios} shows that the two-planet systems with the most distant companions correspond to inner planets with nearly zero eccentricity. The more distant a planet is from its host star, the easier it is to directly image. However, because the outer planets have \textit{no effect} on the inner planets' eccentricities, we cannot obtain any information about the systems' multiplicities; we have no way to determine if a zero-eccentricity RV planet lives alone or has a distant companion. Therefore if one had to choose between directly imaging moderately-eccentric RV planets (where we can quantify the probability that it has an outer companion) or zero-eccentricity planets, we recommend observing the moderately-eccentric planets. We cannot rule out the possibility that the RV population consists of a separate population of zero-eccentricity planets in single-planet systems. We do not entirely discourage direct imaging of zero-eccentricity planets, but the statistically favorable choice is to observe moderately-eccentric planets (see Table 2). 

\subsection{The Uniqueness Problem}
Our simulations and results make interesting predictions, but the distributions used to fit the observed data are not unique. It is possible that other models with different parameters and assumptions could similarly reproduce the observations. Nonetheless we feel our results are important and worth further investigation because we have approached the problem from observational, theoretical, and physical standpoints. The distribution functions for semi-major axis and mass were obtained from statistical analysis of observed RV data. The orbital ranges and masses were determined from both observational data and theoretical assumptions. These were: in stable relaxed planetary systems, planets should orbit outside 2:1 resonances; and if a planetary system has two planets, the outer should be at least as massive as the inner \citep{wright}. Furthermore, with time the number of multiple-planet systems (and systems with long-term trends) discovered by RV has continued to increase, hinting that multiple-planet systems may be more common than the current statistics show. From the physical standpoint, if not all planetary systems have significant planet-planet interactions and scattering, then we might reasonably expect there to be two independent distributions of planetary systems. The planets unaffected by scattering would have $e \sim 0$, while the scattered population of planets would have eccentricities drawn from the Schwarzschild distribution as we have modeled. In this case our two distributions of planets provides a good description of the RV planet population. The predicted frequency of an outer companion provides the key observational test for this conclusion. 

\section{Conclusions}
We have carried out Monte Carlo simulations of mock RV data to determine the statistical effect of undetected wide-separation companions on eccentricity. We quantify this effect and show that, for a given simulated population of planets, the number of eccentric planets increases by $\sim 10\%$ when the outer companion is introduced. Thus undetected outer companions in RV planetary systems can have a small but important effect on the inner planet's Keplerian-fitted eccentricity. We show that the observed RV single-planet eccentricity distribution cannot be matched by 100$\%$ zero-eccentricity planets in double-planet systems, but rather an additional population of high-eccentricity planets in single-planet systems is required. This result lends support to the growing evidence that our solar system, with many low-eccentricity planets, is probably not ordinary. It also agrees with the current RV statistics, which show that high-eccentricity planets constitute a significant portion of the RV population. This is supported by our simulations requiring low-eccentricity double-planets to constitute less than half of the total population, with the best-fitting ratio $R = 42_{-8.7}^{+8.2}\%$. 

Assuming our two simulated populations of planets are a good approximation for the true RV population, we can make predictions about RV planet multiplicity using eccentricity. To do this we calculate the probability that a known RV planet's eccentricity has been biased by a wide-separation companion. Averaged over eccentricity this probability is $\sim 4\%$. For $0.1 < e < 0.3$ and $0.1 < e < 0.2$, the probability is 13$\%$ and 19$\%$, respectively. To test our predictions about the true RV population of planets, RV exoplanet multiplicity, eccentricity as an indicator of multiplicity, and to look for previously undiscovered outer companions, we encourage both high-contrast direct imaging and RV follow-up surveys of known RV single-planets with moderate eccentricities. \\ 
~\\

We would like to thank Matthew Kenworthy for helpful suggestions, comments, and his careful proofreading of this manuscript. We would also like to thank Jared Males and Andy Skemer for helpful insights into radial velocity methods and statistical analysis. We thank the anonymous referee for very helpful comments and suggestions which greatly improved this paper. This work has been supported by the NASA Astrobiology Institute under Cooperative Agreement No.
CAN-02-OSS-02 issued through the Office of Space Science.

\bibliographystyle{aa}
\bibliography{EccPaper}

\newpage
\centering
\begin{table}[t]
\begin{center}
\begin{center}
TABLE 1 \\
Summary of simulation parameters and results
\label{table}
\end{center}
\begin{tabular}{c c}
\hline
\hline
Number of planetary systems ($N$) & 227 \\
Number of simulations & 100 \\
$m_{inner}$  & [0.01,25] $M_{J}$ \\
$m_{outer}$  & [$m_{inner},25~M_{J}$] \\
$a_{inner}$  & [0.05,3] AU \\
$a_{outer}$  & [7,15] AU \\
$e_{single}$ & [0,1] \\
$e_{double}$ & 0 \\
Ratio of double-planet systems to total systems ($R$) & 45$\%$ \\
$\chi_{reduced}^{2}$ & 1.74 \\
Interpolated $R_{min}$ & $42_{-8.7}^{+8.2}\%$ \\
Interpolated minimum $\chi_{reduced}^{2}$ & 1.6 \\
\hline
\hline
\end{tabular}
\end{center}
\end{table}
\clearpage

\newpage
\centering
\begin{table}[t]
\begin{center}
TABLE 2 \\
RV follow-up/direct imaging observing strategy 
\begin{tabular}{|c|c|c|c|}
\hline
RV planet's eccentricity & Comments & RV Follow-up & Direct Imaging \\
\hline
& eccentricity yields no &  &  \\
$e < 0.1$ & information on likelihood & NO & ? \\
 & of outer companion & & \\ 
\hline
 & $\sim 10$-20$\%$ chance of & & \\
$0.1 < e < 0.4$ & having an outer & YES & YES \\
 & companion & & \\ 
\hline
 & extremely unlikely & & \\
$0.4 < e < 1$ & to have outer & NO & NO \\
 & companion & & \\
\hline
\end{tabular}
\end{center}
\end{table}
\clearpage

\end{document}